\documentclass{rsproca}

\jname{rspa}
\Journal{Proc R Soc A\ }

\usepackage{graphicx}
\usepackage[latin1]{inputenc}
\usepackage[USenglish]{babel}
\usepackage[T1]{fontenc}
\usepackage{amsmath,amssymb,amsfonts,stmaryrd,amsthm}
\usepackage{multirow,tabularx,booktabs,ctable}
\usepackage{subfig}
\usepackage{algorithm,algorithmic}

\floatname{algorithm}{\textnormal{Algorithm}}
\restylefloat{algorithm}

\newcommand{\algorithmname}{Algorithm}

\graphicspath{{./Figures/}}

\begin{document}

\title{Changing Computing Paradigms Towards Power Efficiency}

\author{Pavel Klav\'ik$^{1,\ast}$, A.\ Cristiano I.\ Malossi$^{2}$, Costas Bekas$^{2}$, Alessandro Curioni$^{2}$}

\address{$^{1}$Charles University in Prague,\\Faculty of Mathematics and Physics,\\Computer Science
Institute,\\
Malostransk\'e n\'am. 25, 118 00 Prague, Czech Republic\\ \email{klavik@iuuk.mff.cuni.cz}\\ $^{2}$IBM Research - Zurich,\\
S\"aumerstrasse 4, CH-8803, R\"uschlikon, Switzerland \email{\{acm,bek,cur\}@zurich.ibm.com}}

\subject{Computer Science; High Performance Computing; Numerical Analysis}


\corres{Costas Bekas\\
\email{bek@zurich.ibm.com}\\
\small $^\ast$ This work was completed during an internship of the first author at the Computational
Sciences Group, IBM Research - Zurich.}

\begin{abstract}
Power awareness is fast becoming immensely important in computing, ranging from the traditional High Performance
Computing applications, to the new generation of data centric workloads.  

In this work we describe our efforts towards  a power efficient computing paradigm that combines low
precision and high precision arithmetic. We showcase our ideas for the widely used kernel of solving
systems of linear equations that finds numerous applications in scientific and engineering disciplines as well
as in large scale data analytics, statistics  and machine learning.

Towards this goal we developed tools for the seamless power profiling of applications at a fine grain level.
In addition, we verify here previous work on post FLOPS/Watt metrics and show that these can shed much more light
in the power/energy profile of important applications.

\end{abstract}

\maketitle

\section{Introduction} \label{sec:introduction}

Since the 1970's the High Performance Computing community, including industry and academia, have
tremendously benefited from the existence of a single, simple, and easy to deploy benchmark: the
LINPACK benchmark~\cite{Dongarra2003}. Its original target was the solution of a dense
linear system of 1000 equations in the minimum time, thus generating a ranking of the most powerful
supercomputers based on the measured number of floating point operations per unit of time (FLOPS).
As supercomputers became increasingly more potent, the LINPACK benchmark evolved allowing to scale
the problem size, while keeping the same goal. Indeed, while early machines achieved Mega-FLOPS,
today's most powerful supercomputers reach Peta-FLOPS, which constitutes an astonishing increase of
nine orders of magnitude in less than 35~years.

Since 1993 a list of the most powerful supercomputers has been kept and updated bi-annually in June
and November: this is the famous TOP500~list\footnote{\url{http://top500.org}}. It graphically
portraits this exponential increase in performance, that is attributed mostly to Moore's law, which
in turn has been widely (and vaguely) interpreted as an exponential performance improvement in CMOS
technology. However, during the last decade, the core technology has dramatically changed. Voltage
and thus frequency scaling has stopped, pointing out the true nature of Moore's law, which
essentially predicted our technological capacity to exponentially increase the number of logical
gates on a chip. To continue the exponential overall improvements, technologists have turned into
multi-core chips and parallelism at all scales. However, this new trend has lead to a series of new
issues and questions, which nowadays represent the most critical aspects in the development of
future architectures and software. These questions are mainly: \emph{(i)}~How to program such complex
machines? \emph{(ii)}~How to cope with the very fast increase in power requirements? In this work we
concentrate on the latter one, although we believe that programmability and energy efficiency will
ultimately prove to be closely connected.

In~\cite{BekasCurioni2010}, the authors have shown that very important power savings can be achieved by
a careful combination of different precision levels for floating point arithmetic. Those
observations led to introduction of new energy-aware performance metrics. This was motivated by
the fact that the traditional energy-aware performance metrics (derived from the classic FLOPS
performance metric) have serious shortcomings and can potentially draw a picture that is far from
reality. For instance, the widespread FLOPS/Watt metric, which is used to rank supercomputers in
the Green500~list\footnote{\url{http://green500.org}}, might still promote power hungry algorithms
instead of other more energy efficient techniques. In contrast, by using metrics based on both time
to solution and energy, as proposed in~\cite{BekasCurioni2010}, it is possible to get a much more
realistic picture, which can drive and help the design of future energy efficient machines and
algorithms. At the same time, we also need to abandon the (convenient and simple) LINPACK benchmark
and make use of a relatively small selected set of benchmarks, characterized by widely different
features and thus representative of the majority of the current
applications~\cite{Asanovic2006,Kaltofen2012}.

The reasoning for the mixed precision computing paradigm and the associated energy-aware performance
metrics  in~\cite{BekasCurioni2010} is based on a model of power consumption for the various
subsystems of a computing platform, such as the floating point unit, the interconnect, and the
memory; this model is an extrapolation of future exascale machines~\cite{Kogge2009}. At the same
time, it is widely acknowledged that actual on-line measurements can be complicated  and require
external circuits that cast noise in the measurements. Recent advances in chip technology however
have allowed the on chip measurement of instantaneous power consumption by means of integrated
sensors~\cite{Brochard2010,FloydEtAll2010,Kephart2007,Lefurgy2007}. In this work we advance the
initial contribution in~\cite{BekasCurioni2010} and describe a fully automated, on-line, minimally
invasive system for measuring power performance. Our tool follows a similar philosophy to other tools such as 
PowerPack~3.0\footnote{\url{http://scape.cs.vt.edu/software/powerpack-3-0/}}
package~\cite{powerpack}, however we have put here emphasis on minimizing measurement noise and fine granularity.  
Our system allows users to easily instrument their code, as well as to
directly distinguish and separately measure power performance of various code segments.
The tool is completed by a carefully designed graphical interface, which provides real-time,
easy-to-read informative plots and diagrams.

The target platform of our analysis is the IBM~POWER7 processor, that offers a wide variety of
on-chip sensors.  On this machine we perform a series of analysis that confirms and extends findings
and theories devised in~\cite{BekasCurioni2010} on an IBM~Blue~Gene/P system.  More importantly, we
outline a general energy-aware framework for solving linear systems, that is based on a careful
combination of mixed-precision arithmetics. In addition, we take a further step and show that our
framework allows the use of inaccurate data, as well as inaccurate computations. That is
particularly crucial as data movement is expected to completely dominate running time and power
consumption in future systems.

This work is organized as follows. In Section~\ref{sec:experimental_setting} we describe our
framework for on-line, on chip power consumption measurements.
In Section~\ref{sec:theoretical_background} we describe the inexact computing/data framework for
solving systems of linear equations. Then, in Section~\ref{sec:experimental_results} we present our
experimental results applied to covariance matrices from applications in data-analytics. We conclude
with remarks in Section~\ref{sec:conclusions}.


\section{On-line -- on chip power measurement} \label{sec:experimental_setting}

\begin{figure}[t!]
\centering
\includegraphics[scale=0.8]{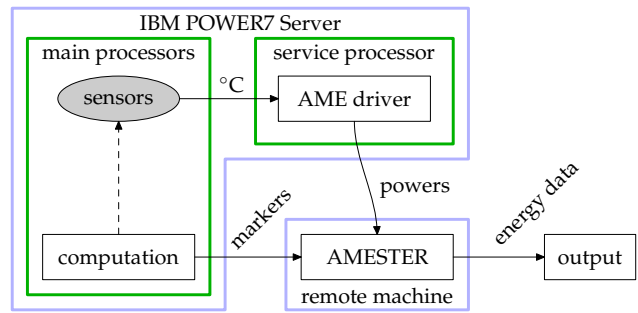}
\caption{Schematic description of the experimental framework for power measurements.}
\label{pic:amester_scheme}
\end{figure}

In this section we describe the framework that we set up to perform power measurements. Our system
consists of an IBM~blade server~PS702\footnote{%
\url{http://www-03.ibm.com/systems/bladecenter/hardware/servers/ps700series/specs.html}} \cite{IBMRedBookPS702},
with two IBM~POWER7 chips (octa~core, 3.00~GHz CPU and 4~MB~L3-cache each) and 64~GB of system
memory. Each POWER7~chip has several embedded thermal sensors which measure the temperature of the
different components. Data from sensors are transferred to a separate service processor that is
embedded in the POWER7 server. The service processor is running a firmware called AME Driver which
in turn calculates power from the thermal sensors as described
in~\cite{FloydEtAll2010,Kephart2007,Lefurgy2007}. We remark that it is crucial that power readings
are calculated on the service processor, so they do not add background noise to the actual
user computation; in this regard the POWER7 server is an ideal platform to perform power
measurements. Finally, on a remote machine we run a software named AMESTER (Automated Measurement of
Systems for Energy and Temperature Reporting). This is an experimental application developed by IBM,
which periodically retrieves power measures from the service processor of the POWER7 server and
collects them into a file (for more details see~\cite{Kephart2007,Lefurgy2007}). A graphical
description of the power measurement framework is depicted in \figurename~\ref{pic:amester_scheme}.

To ease the instrumentation of the code, we have developed a library that provides a black-box
interface between AMESTER and the running application. This layer uses TCP/IP sockets hidden behind
a series of simple calls. Issues of TCP/IP latency and resolution of instrumentation are directly
resolvable since the AME Driver allows for a store/trace forward mode, where a bunch of power data
is measured and collected such that it can be subsequently read remotely. Thus instrumentation
permits to setup and run AMESTER, and then mark power data during computation; using this framework we
can map measured data to different parts of the code and get a clear picture of the specific power
consumption for each analyzed kernel in our application. The resolution of power is less than 1~W
while the sampling frequency is approximately~1~kHz (i.e., up to 1000~samples per second).
The error of the power measurements is stated to be
lower than 2\% \cite{Knobloch2013,FloydEtAll2010,IBMRedBookPS702}.

\begin{figure}[t!]
\centering
\includegraphics[width=0.5\textwidth]{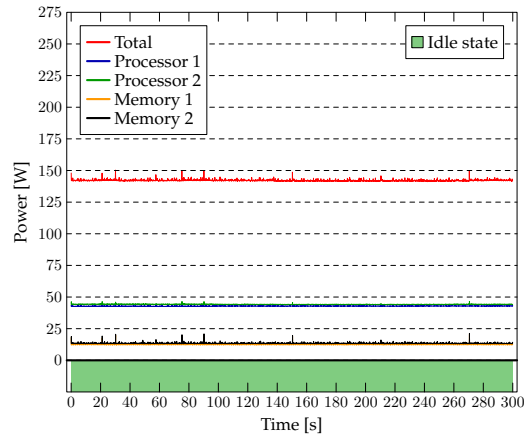}
\caption{Power of chips and memories on the IBM blade server PS702 during an idle state.}
\label{pic:empty_machine}
\end{figure}

\begin{table}[t!]
\centering
\caption{Average power (over 300 seconds) of chips and memories on the IBM blade server PS702 during
an idle state. }
\begin{tabular}{lcc}
\toprule
Sensor      & Average power [W] & Standard error [W]\tabularnewline
\midrule
Total       &           142.1   & 0.6 \tabularnewline
Processor 1 & \phantom{0}42.8   & 0.2 \tabularnewline
Processor 2 & \phantom{0}44.2   & 0.2 \tabularnewline
Memory 1    & \phantom{0}12.5   & 0.4 \tabularnewline
Memory 2    & \phantom{0}13.4   & 0.4 \tabularnewline
\bottomrule
\end{tabular}
\label{tab:average_power_idle}
\end{table}

An example of power analysis is illustrated in Figure~\ref{pic:empty_machine}, where we plot
measurements from five power sensors during an idle state of the machine. From top
to bottom these are: total power of the system, power of each of the two processors, and power of
each of the corresponding memory banks. Note that the total power of the system includes additional
sources of power consumption, such as disks and fans, which in the idle state consume approximately~30~W.
These values are averaged in time in \tablename~\ref{tab:average_power_idle},
providing us the lower bound limit of power consumption of the PS702~blade~server.
In the examples presented in the forthcoming sections, we will observe that chip
power can reach up to 80~W, while memory power can get up to 40~W. Total power is thus always between~140~W and~200~W.

\section{Inexact computing for the solution of dense linear systems} \label{sec:theoretical_background}

\newcommand{\norm}[1]{\left\lVert#1\right\rVert}
\newcommand{\transpose}{\mathsf{T}}

Let us focus on the solution of a dense system of $n$~linear equations in the form
\begin{equation}
\mathrm{A} \boldsymbol x  =  \boldsymbol b,
\label{eq:problemSingleRHS}
\end{equation}
where $\mathrm{A} \in \mathbb{R}^{n \times n}$ is a symmetric and positive definite matrix,
$\boldsymbol x\in \mathbb{R}^{n}$ the solution vector, and $\boldsymbol b \in \mathbb{R}^{n}$ the
right-hand side. This problem can be generalized to the case of $m$~right-hand sides $\boldsymbol
b_1,\dots,\boldsymbol b_m$ such that we can write
\begin{equation}
\mathrm{A} \mathrm{X}  =  \mathrm{B},
\label{eq:problemMultipleRHS}
\end{equation}
where $\mathrm{B} = \left[\boldsymbol b_1, \boldsymbol b_2, \ldots, \boldsymbol b_m\right] \in
\mathbb{R}^{n \times m}$ and $\mathrm{X} = \left[\boldsymbol x_1, \boldsymbol x_2, \ldots,
\boldsymbol x_m\right] \in \mathbb{R}^{n \times m}$ are two rectangular matrices containing all the
right-hand sides and solution vectors, respectively.

The target of our analysis is the solution of large size covariance matrix problems, which are
crucial for uncertainty quantification and applications in data-analytics (see, e.g.,
\cite{BekasCurioniFedulova2009} and references therein). With this aim, in the following we describe
two iterative refinement procedures derived from the Cholesky and the conjugate gradient~(CG)
methods for the solution of this class of problems.

\subsection{Cholesky with iterative refinement}\label{subsec:CholeskyIterativeRefinement}

The Cholesky decomposition $\mathrm{A} = \mathrm{R}^\transpose \mathrm{R}$~(see,
e.g.,~\cite{GolubVanLoan1996}) offers a cache friendly and high performance approach for the
solution of problems~\eqref{eq:problemSingleRHS} and~\eqref{eq:problemMultipleRHS}.  The overall
complexity of the decomposition is $1/3\,n^3 + \mathcal{O}(n^2)$.  Once it is done, the solution of
the problem can be computed very fast, even for multiple right-hand sides:
\begin{equation}
\mathrm{X} = \mathrm{R}^{-1} \mathrm{R}^{-\transpose} \mathrm{B},
\label{cholesky}
\end{equation}
where we remark that $\mathrm{R}$ is an upper-triangular matrix; consequently the cost
of~\eqref{cholesky} is~$\mathcal{O}(n^2)$ per right-hand side vector, thus in general negligible.

In~\cite{BekasCurioniFedulova2009,BekasCurioniPatent2012} (see also references therein), it has been
shown that to cope with the formidable cubic cost of the Cholesky decomposition, mixed precision
iterative refinement can be employed. For instance, let us consider the procedure described in
\algorithmname~\ref{algo:CholeskyIterativeRefinement} for the case of a single right-hand side,
where we use the~``\textasciitilde''~symbol to denote matrices or vectors stored in low precision.
\begin{algorithm}[!t]
\algsetup{indent=2em}
\caption{: Cholesky method with iterative refinement for one right-hand side.}
\label{algo:CholeskyIterativeRefinement}
\begin{algorithmic}[1]
\STATE Compute  Cholesky decomposition $\mathrm{A} \approx \tilde{\mathrm{R}}^\transpose
		\tilde{\mathrm{R}}$.  \hfill\textbf{[Low precision]} 
\STATE Generate approximate solution ${\boldsymbol x} = \tilde{\mathrm{R}}^{-1}
		\tilde{R}^{-\transpose } \boldsymbol b$. \hfill\textbf{[Low precision]} 
\STATE Compute residual $\boldsymbol r= \boldsymbol b-\mathrm{A} \boldsymbol x$. \hfill\textbf{[High precision]} 
\WHILE{$\norm{\boldsymbol r}_2  \ge \text{tolerance}$}
    \STATE \label{algo:CholeskyIterativeRefinement_Error} Compute error components ${\boldsymbol z}
	= \tilde{\mathrm{R}}^{-1} \tilde{\mathrm{R}}^{-\transpose } \boldsymbol r$. \hfill\textbf{[Low precision]} 
    \STATE Update solution $\boldsymbol x=\boldsymbol x+{\boldsymbol z}$. \hfill\textbf{[High precision]} 
    \STATE Compute residual $\boldsymbol r= \boldsymbol b-\mathrm{A} \boldsymbol x$. \hfill\textbf{[High precision]} \hspace{0em}
\ENDWHILE
\end{algorithmic}
\end{algorithm}
%
%
On modern multicore chips with accelerators, the computation of the Cholesky decomposition at the
very first step can be speedup by using IEEE~single precision arithmetics. The resulting
inexact-decomposition is then used also at line~\ref{algo:CholeskyIterativeRefinement_Error} of
\algorithmname~\ref{algo:CholeskyIterativeRefinement} for each iteration till convergence.

Observe however that the complexity of the overall scheme remains~$\mathcal{O}(n^3)$, due to the
matrix factorization. Therefore, even with significant single precision acceleration,
time-to-solution increases cubically with the matrix size and dominates the computation in case of
large problems.  A second important remark concerns the very nature of dense matrix factorizations,
such as the Cholesky decomposition. These are rich in BLAS-3, cache friendly linear algebra
primitives, a property that has been the focus of attention for optimizing system performance. It
reflects that codes like the Cholesky decomposition tend to make full use of the computational
resources of the system, stressing its components to their limits.

\begin{figure}[t!]
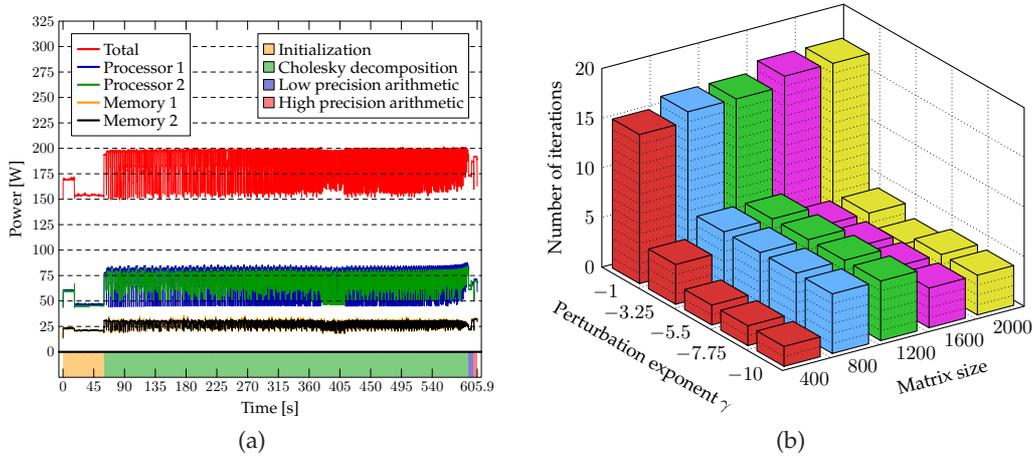

\centering
\subfloat[\label{pic:cholesky_example}]{\includegraphics[width=0.475\textwidth]{energy_paper2_chol_example.eps}}\hfill
\subfloat[\label{pic:cholesky_perturbation}]{\includegraphics[width=0.475\textwidth]{energy_paper19_cholesky_iterations.eps}}
\caption{(a) Power diagram on IBM~PS702~Blade Server during the solution of a dense linear system
with 32~right-hand sides, employing the Cholesky based approach with iterative refinement. In the
timeline, each computational phase is displayed with a different color.\\(b) The number of
iterations to solution of a perturbed Cholesky algorithm, for a range of matrix sizes and
perturbation exponents.}
\end{figure}

As an example, \figurename~\ref{pic:cholesky_example} shows power measurement for solving a linear system
with 32~right-hand sides using the Cholesky decomposition and mixed precision iterative refinement.
It is evident that the majority of time and energy is spent for the Cholesky decomposition.
In addition, note that power consumptions peaks around 200~W, due to the BLAS-3 primitives that ensure
high sustained performance in the arithmetic computations.

The burning question thus, is whether it is possible to fundamentally reduce the computational cost.
Iterative refinement theory predicts that the only requirement for the process to converge is that
the solver used for the inner step at line~\ref{algo:CholeskyIterativeRefinement_Error}
of \algorithmname~\ref{algo:CholeskyIterativeRefinement} is not too ill-conditioned~\cite{Higham.96}.
In particular, the following conditions need to be satisfied:
\begin{equation}
\label{eq:iterref}
(\mathrm{A}+\Delta \mathrm{A}) {\boldsymbol z} = \boldsymbol  r \quad \textrm{ and } \quad
\|\mathrm{A}^{-1} \Delta \mathrm{A}\|_\infty <1,
\end{equation}
where $\Delta \mathrm{A}$~is a given perturbation of the matrix coefficients. In
the iterative refinement setting, this perturbation corresponds to the error introduced by storing
the matrix (or its factorization), in low precision.  Bound~\eqref{eq:iterref} implies that if the
problem is not too ill-conditioned, then the iterative scheme would still be able to converge with
the desired high precision, even if the accuracy of the internal solver is quite low.


To see this, consider the results of~\cite{BekasCurioniFedulova2009,BekasCurioniPatent2012}
depicted  in~\figurename~\ref{pic:cholesky_perturbation}: there we plot the
number of iterations to solve a dense linear system up to working precision (in our case,
IEEE~double precision), by using a perturbed Cholesky algorithm. In particular, the
factor~$\mathrm{R}$ (computed in double precision) is perturbed with a range of upper-triangular
matrices~$\mathrm{E}$, whose norm is such that $\|\mathrm{R}+\mathrm{E}\|_2/\|\mathrm{R}\|_2 \le
10^\gamma$, with $\gamma \in
\{-1,-3.25,-5.5,-7.75,-10\}$.  This experiment evidently demonstrates that even in presence of large
perturbations, e.g.,~$\gamma = -1$, the method is still able to converge. Moreover, even if the
number of iterations increases, each one of them costs $\mathcal{O}(n^2)$ which is generally several
orders of magnitude less than the Cholesky decomposition.


This example shows clearly that the accuracy of the employed solver can be lowered without
compromising the robustness of the numerical scheme.  In fact, the very nature of the solver we
choose is quite irrelevant, with the only crucial property being that of bound~\eqref{eq:iterref}.

\subsection{Conjugate gradient with iterative refinement}

In~\cite{BekasCurioniFedulova2009} (see also~\cite{BekasCurioniPatent2012}) the conjugate gradient
method with iterative refinement (see~\algorithmname~\ref{algo:CGIterativeRefinement}) is proposed
as a possible energy-efficient replacement of the Cholesky based approach.  Particularly, it is
shown that the overall computational cost can be reduced down to quadratic levels from the original
cubic cost of dense solvers.

\begin{algorithm}[!t]
\algsetup{indent=2em}
\caption{: Conjugate gradient method with iterative refinement for one right-hand side.}
\label{algo:CGIterativeRefinement}
\begin{algorithmic}[1]
\STATE Set initial solution $\boldsymbol x = {\boldsymbol x}_0$. 
\STATE Compute residual $\boldsymbol r= \boldsymbol b-\mathrm{A} {\boldsymbol x}$. \hfill\textbf{[High precision]} 
\STATE Set initial direction $\boldsymbol p= \boldsymbol r$. 
\WHILE{$\norm{\boldsymbol r}_2  \ge \text{tolerance}$}
    \STATE Compute matrix-vector multiplication $\boldsymbol z = \tilde{\mathrm{A}} \boldsymbol p$. \hfill\textbf{[Low precision]} 
    \STATE Compute $\rho = {\boldsymbol r^\transpose \boldsymbol r}$ and $\alpha = {\rho}  /
	{\left(\boldsymbol p^\transpose \boldsymbol z\right)}$. \hfill\textbf{[High precision]} 
    \STATE Update solution $\boldsymbol x= \boldsymbol x+\alpha \boldsymbol p$. \hfill\textbf{[High precision]} 
    \STATE Update residual $\boldsymbol r= \boldsymbol r-\alpha \boldsymbol z$. \hfill\textbf{[High precision]} 
    \STATE Compute $\beta =\left(\boldsymbol r^\transpose \boldsymbol z\right) / \rho$ and update
	direction $\boldsymbol p=\boldsymbol r+\beta \boldsymbol p$. \hfill\textbf{[High precision]} \hspace{0em}
\ENDWHILE
\end{algorithmic}

\end{algorithm}

\begin{figure}[!t]
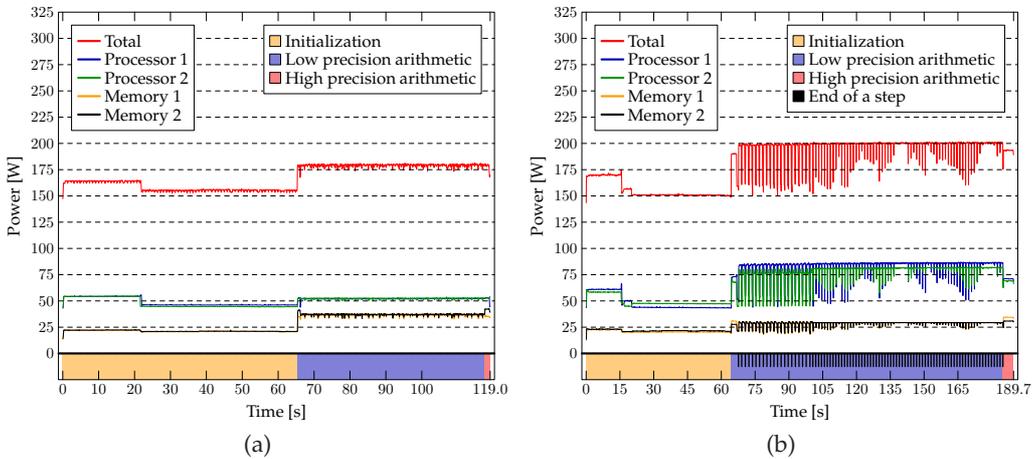

\centering
\subfloat[\label{pic:cg_one_RHS}]{\includegraphics[width=0.49\textwidth]{energy_paper4_cg_one_RHS.eps}}\hfill
\subfloat[\label{pic:cg_32_RHSs}]{\includegraphics[width=0.49\textwidth]{energy_paper5_cg_32_RHSs.eps}}
\caption{Power diagrams on IBM~PS702~Blade Server during the solution of a dense linear system using
the conjugate gradient algorithm with iterative refinement. In the timeline, each computational phase
is displayed with a different color, while the end of each step is marked by a black line. (a)~1
right-hand side. (b)~32 right-hand sides.}
\label{pic:cg_RHS}
\end{figure}

\figurename~\ref{pic:cg_one_RHS} illustrates power measurements for this approach, solving the same
problem as in the Cholesky based version, but with one right-hand side. First of all, we observe that
time-to-solution is strongly reduced as compared to the Cholesky based method. Indeed, there is no initial
factorization phase any longer that dominates the run time.  A second observation regards the peak power
consumption which, with respect to the idle state, is about 50\% less than the
peak power observed for the Cholesky case in~\figurename~\ref{pic:cholesky_example}.

When increasing the number of right-hand sides from~1 to~32, as shown
in~\figurename~\ref{pic:cg_32_RHSs}, the major internal kernel of the CG based algorithm becomes the
matrix-matrix multiplication, and thus peak power characteristics resemble those of the Cholesky
based approach.  The sudden decreases in power are due to executing vector operations (BLAS-2),
which do not stress the system as much as BLAS-3 based computations. We also notice that the first
few steps take longer time before the machine moves the data in right caches.  As a consequence,
in the case of the multiple right-hand sides, it is by far the decreased time-to-solution that
drives the decrease in total energy of the CG based method with respect to the Cholesky one.


\section{Iterative refinement on approximate covariance matrices} \label{sec:experimental_results}

Bound~\eqref{eq:iterref} offers a second reading: we can cast the error not to the employed solver, but to the data.
Indeed, this is a statement akin to the classical backward stability. In other words,
we can use an~\emph{exact}~solver that works on~\emph{inexact}~data, as shown in the example of
the perturbed Cholesky factor discussed in Section~\ref{sec:theoretical_background}.

A possible strategy to take advantage of this approach, is to approximate the system matrix
by following the pattern in its structure.
For instance, in several applications matrices exhibit a progressive decay of their elements away from the main diagonal.
In those cases, a reduced banded version of the matrix (with just a few diagonals) can be used, in place of the complete one.
Observe that the resulting complexity of the inner solver is drastically reduced, as well as data movement requirements, which are lowered by almost one order of magnitude.

In general, we can opt for sampled versions of the matrices involved for the inner solver. To this
end we can borrow approaches developed in the literature of randomized linear algebra (see, e.g.,
\cite{tropp.siamrev.randomized} and references therein). The main idea is that matrix~$\mathrm{A}$ is
replaced by $\tilde{\mathrm{A}} = \mathrm{P}\mathrm{A}\mathrm{P}^\transpose$ where $\mathrm{P}$~is a suitably chosen fat sketching matrix $\mathrm{P} \in
\mathbb{R}^{k \times n}$ where $k \ll n$.

In the following, we showcase the validity of our arguments with a set of numerical experiments using model covariance
matrices with controlled degree of decay~$d$ away from the main diagonal:
\begin{equation*}
\mathrm{A}(i,j) =
\left\{
\begin{array}{ll}
\dfrac{1}{|i-j|^d} & i \ne j,\\[2ex]
1+\sqrt{i} & i=j.
\end{array}
\right.
\end{equation*}
%
In all our experiments, the right-hand side vectors are generated randomly
with coefficients in~$[0,1]$ while the tolerance for the stopping criteria is set equal to~$10^{-5}$.
Note that matrices with different degree of decay are equally difficult for the standard Cholesky
based solver. In contrast, for iterative solvers, such as the CG based method, matrices that exhibit
strong decay (i.e., high~$d$) are easier to solve with respect to those with a small or null decay,
since they become strongly diagonally dominant.

\subsection{Time and energy analysis}

\begin{figure}[!t]
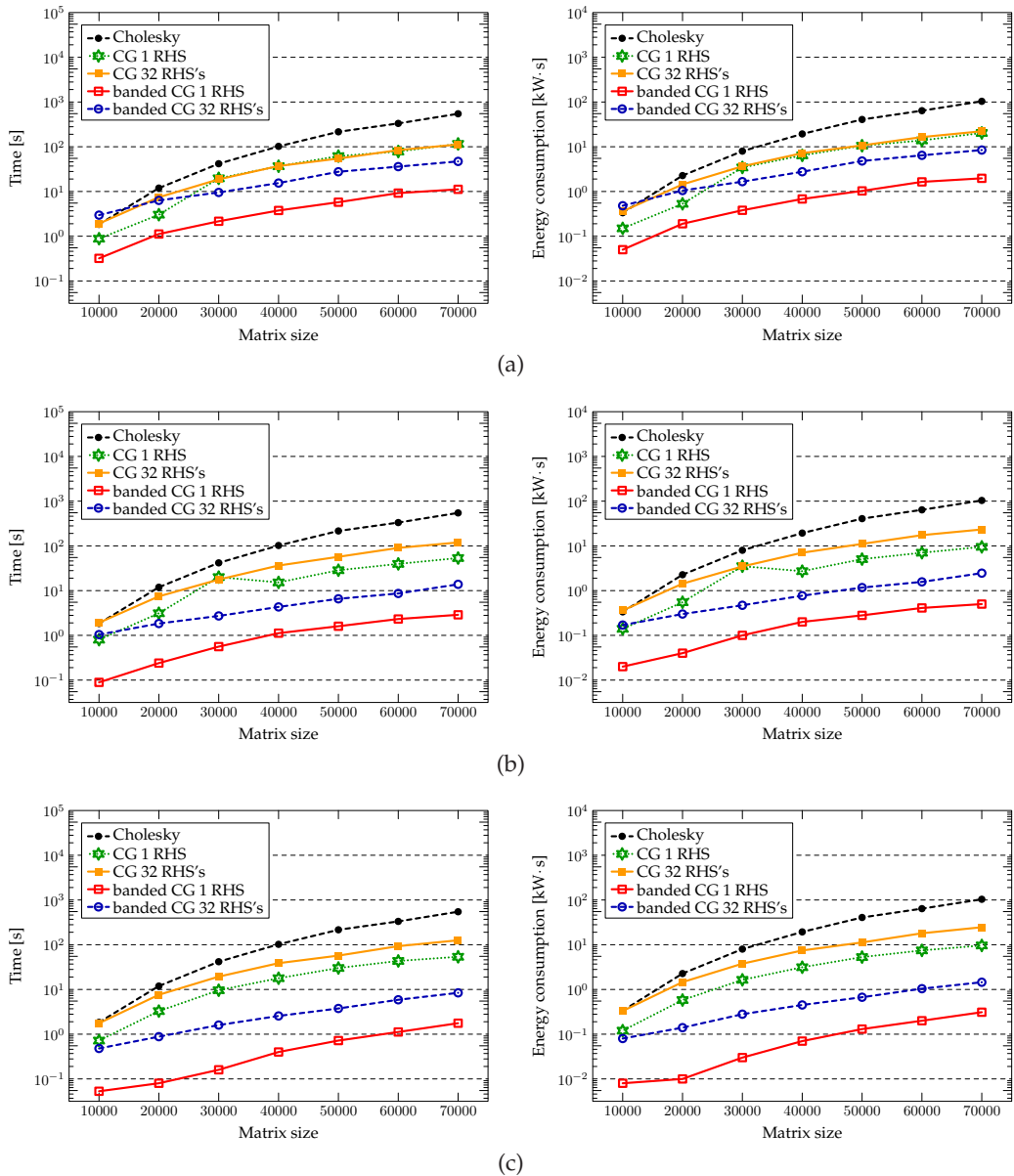

\centering
\subfloat[\label{pic:comparison_power1}]{\includegraphics[width=1\textwidth]{energy_paper8_comparism_power1.eps}}\\[2ex]
\subfloat[\label{pic:comparison_power2}]{\includegraphics[width=1\textwidth]{energy_paper9_comparism_power2.eps}}\\[2ex]
\subfloat[\label{pic:comparison_power4}]{\includegraphics[width=1\textwidth]{energy_paper10_comparism_power4.eps}}
\caption{Time and energy comparison between different methods for the solution of power model
covariance matrices of different size. (a)~$d=1$. (b)~$d=2$. (c)~$d=4$.}
\label{pic:comparison_power}
\end{figure}

\figurename~\ref{pic:comparison_power} illustrates time-to-solution and energy
consumption to solve problem~\eqref{eq:problemSingleRHS} or~\eqref{eq:problemMultipleRHS}
with different iterative refinement based methods and for a range of complete and approximate covariance matrices, with $d=[1,2,4]$.
All the quantities are calculated without accounting for the initialization phase.
The Cholesky based method shows no difference with respect to the degree of
decay, since it has a fixed complexity that depends only on the matrix size.  A similar result is also obtained for the non-banded CG based method.
However, by using the banded CG version, with a bandwidth of~33~(16 on each side of the main
diagonal), we observe a sensible progressive reduction in the time-to-solution, and thus in the
total consumed energy.


\begin{table}[!t]
\centering
\caption{Comparison of different methods to solve a model covariance matrix, with $n = 70000$.}
\scalebox{0.93}{\setlength{\tabcolsep}{4pt}%
\begin{tabular}{lccccccccc}
\toprule
\multirow{2}{*}{} & \multirow{2}{*}{RHS's} & Iterations & \multirow{2}{*}{Time [s]} & Average& Standard & Energy & TOP500 & Green500 \tabularnewline
     & & [low/high] & & power [W] & error [W] & [kW$\cdot$s] & [GFlops] & [GFlops/W]\tabularnewline
\toprule
$d=1$\tabularnewline
\midrule
Cholesky & 32 & \phantom{00}1 / 1 & 546.0& 190.0  & 13.5 & 103.7  & 214.4& 1.11 \tabularnewline
CG &1  & 179 / 2  & 115.7& 175.7  &  \phantom{0}3.2 &  \phantom{0}20.3  & \phantom{0}15.3 & 0.09 \tabularnewline
CG &32 & \phantom{0}83 / 1 & 112.5& 197.1  &  \phantom{0}7.5 &  \phantom{0}22.2  & 233.8& 1.19 \tabularnewline
Banded CG &1  & 691 / 8 & \phantom{0}11.2& 175.9  &  \phantom{0}6.2 &   \phantom{00}2.0  & \phantom{00}7.1  & 0.04 \tabularnewline
Banded CG &32 & 692 / 8 & \phantom{0}47.0& 179.4  & 15.3 &   \phantom{00}8.4  & \phantom{0}54.3 & 0.30 \tabularnewline
\toprule
$d=2$\tabularnewline
\midrule
Cholesky & 32 & \phantom{00}1 / 1 & 546.0& 190.0  & 13.5 & 103.7  & 214.4& 1.11 \tabularnewline
CG &1         & \phantom{0}83 / 1 & \phantom{0}53.5& 177.8  &  \phantom{0}3.1 &  \phantom{00}9.5  & \phantom{0}15.3 & 0.09 \tabularnewline
CG &32        & \phantom{0}85 / 1 & 119.3& 195.2  & 11.2 &  \phantom{0}23.3  & 225.9& 1.16 \tabularnewline
Banded CG &1  & 173 / 2 & \phantom{00}2.9& 174.4  &  \phantom{0}7.0 &   \phantom{00}0.5 & \phantom{00}6.3 & 0.04 \tabularnewline
Banded CG &32 & 177 / 2 & \phantom{0}13.8& 177.6  & 14.6 & \phantom{00}2.5  & \phantom{0}46.1 & 0.26 \tabularnewline
\toprule
$d=4$\tabularnewline
\midrule
Cholesky & 32 & \phantom{00}1 / 1 & 546.0& 190.0  & 13.5 & 103.7  & 214.4& 1.11 \tabularnewline
CG &1         & \phantom{0}85 / 1 & \phantom{0}53.8& 179.0  &  \phantom{0}1.8 &   \phantom{00}9.6  &  \phantom{0}15.7& 0.09 \tabularnewline
CG &32        & \phantom{0}88 / 1 & 125.5& 195.0  & 10.8 &  \phantom{0}24.6  & 222.2& 1.13 \tabularnewline
Banded CG&1  & \phantom{0}85 / 1 & \phantom{00}1.8& 174.1  & \phantom{0}4.9 &   \phantom{00}0.3  &   \phantom{00}5.5& 0.03 \tabularnewline
Banded CG&32 & \phantom{0}88 / 1 & \phantom{00}8.4& 172.6  & 14.2 &   \phantom{00}1.5  &  \phantom{0}37.8& 0.22 \tabularnewline
\bottomrule
\end{tabular}}
\label{tab:power_table}
\end{table}

To quantify the benefits of the banded CG based method with respect to the other approaches,
in \tablename~\ref{tab:power_table} we report the values obtained for a covariance matrix with $n=70000$.
In most of the cases, the banded CG~based version is able to reduce the time-to-solution by a factor ten or more,
with respect to the non-banded version.
This gain is mainly due to the reduced time~cost of each low precision iteration.
This is evident in the case~$d=4$, where the banded CG~based version converges between 15~and 30~times faster than the non-banded version, despite the number of low and high precision iterations is exactly the same for both methods.
In addition the peak power consumption does not increase when solving for more than one right-hand side,
leading to an additional reduction in the total energy consumed by the problem.
For instance, in the case of~$d=4$ and~$n=70000$, the energy consumed by the banded CG version with~32 right-hand sides is
just~6\% of the energy consumed by the non-banded CG version to solve exactly the same problem.

\subsection{Standard performance metrics analysis}

Despite the energy analysis in the previous section highlighted the strong potential of the inexact iterative refinement strategy,
by ranking the methods with standard metrics, i.e., the TOP500~[GFlops] and the Green500~[GFlops/W], we obtain a surprising result:
 the banded CG based method scores quite poorly in both rankings (see \tablename~\ref{tab:power_table}, rightmost columns).

\begin{figure}[!t]
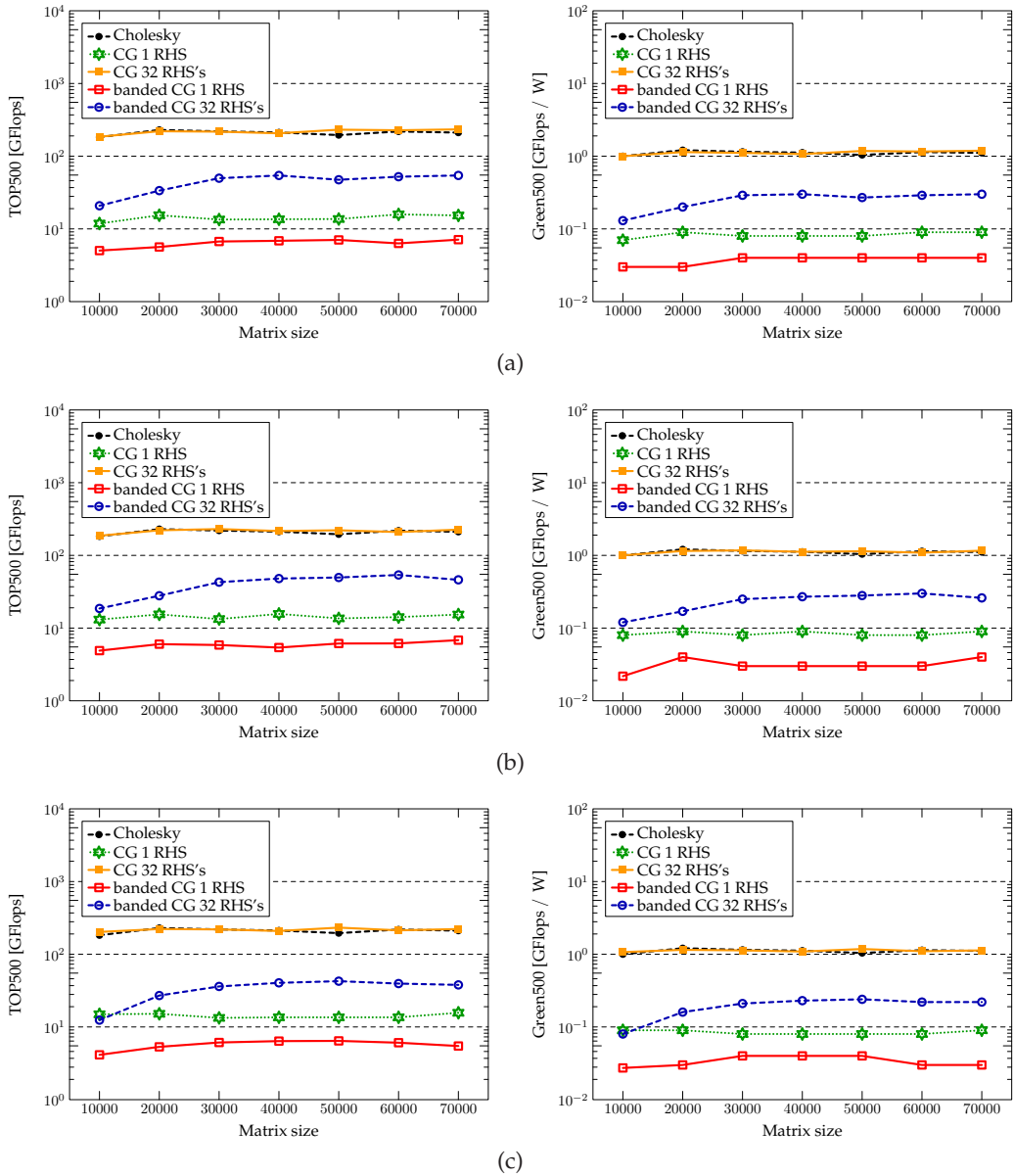

\centering
\subfloat[\label{pic:standard_comparison_power1}]{\includegraphics[width=1\textwidth]{energy_paper12_standard_comparison_power1.eps}}\\[2ex]
\subfloat[\label{pic:standard_comparison_power2}]{\includegraphics[width=1\textwidth]{energy_paper13_standard_comparison_power2.eps}}\\[2ex]
\subfloat[\label{pic:standard_comparison_power4}]{\includegraphics[width=1\textwidth]{energy_paper14_standard_comparison_power4.eps}}
\caption{Standard metrics applied to a range of methods for the solution of power model covariance matrices of different size.
 (a)~$d=1$. (b)~$d=2$. (c)~$d=4$.}
\label{pic:standard_comparison_power}
\end{figure}

This behavior is also confirmed in \figurename~\ref{pic:standard_comparison_power}, for a large range of matrix sizes.
The motivations behind these results are linked to the nature of the employed methods. On the one hand,
the Cholesky and the non-banded CG (especially with multiple right-hand sides) based methods
make extensive use of computational resources, performing a very large number of operations per second
to factorize the matrix and to perform dense matrix-vector multiplications, respectively. On the other hand,
the non-banded CG based method performs much less multiplications per second,
due to the reduced effective size of the matrix, which is also stored in a different
format.
Therefore, the number of BLAS-2 and BLAS-3 operations for the banded CG based method is drastically reduced,
leading to a poor score in the TOP500 ranking.


Regarding energy, we observe that the curves in the right graphs are quite similar to those for the time on the left.
The reason is that average power differs for methods by at most 20~W, so that it has a very low
impact on the energy trend with respect to time-to-solution, which is dominant. This also reflects
in the Green500 ranking, where the result is dominated by the number of GFlops, regardless of the
effective average power consumption, and thus leading to similar conclusions with respect to the
TOP500 ranking.  However, those 20~W actually corresponds to a significant 33\% difference with
respect to the net power consumption (from 142.1~W up to 200~W).

In summary, our results show clearly that standard metrics are not able to capture the real behavior
of applications, and this is mainly due to the fact that they rank methods and machines on the base
of GFlops and GFlops/W rather than real quantities of practical interest, i.e., time-to-solution and
energy.  In this regard, the new strategies proposed in~\cite{BekasCurioni2010} represent a first
step towards the development of concrete energy-aware metrics which, combined with the use of
methods as CG in place of LINPACK, can drive the development of future hardware in the correct
direction.



\section{Conclusions}\label{sec:conclusions}

Power awareness and energy consumption is quickly becoming a central issue in HPC~systems and
research. There is rich recent activity towards solutions that allow us to continue historical
exponential improvements in performance while at the same time keeping power requirements at
affordable levels. Towards this direction, we described here our recent advancements in the context
of energy-aware performance metrics for HPC~systems. More in detail, we developed a framework for
accurate, on chip and on-line power measurements, with minimal distortion of user runs. Our tool
allows the easy instrumentation of user code which subsequently enables its detailed power
profiling. We demonstrated the validity and superiority of our proposed performance metrics on
actual energy measurements. Finally, we described a framework that combines different precision
levels and relaxes data fidelity, and thus has the potential to reduce energy by a few orders of
magnitude.


\section*{Acknowledgments}

The project Exa2Green (under grant agreement n$^\circ$318793) acknowledges the financial support of
the Future and Emerging Technologies (FET) programme within the ICT theme of the Seventh Framework
Programme for Research (FP7/2007-2013) of the European Commission. IBM, ibm.com, and POWER7 are
trademarks or registered trademarks of International Business Machines Corp., registered in many
jurisdictions worldwide. Other product and service names might be trademarks of IBM or other
companies. A current list of IBM trademarks is available on the Web at ``Copyright and trademark
information'' at: \url{www.ibm.com/legal/copytrade.shtml}.

\bibliographystyle{plain}
\bibliography{energy_consumption_paper}

\end{document}